\documentclass[sigconf,nonacm]{acmart}
\AtBeginDocument{%
  \providecommand\BibTeX{{%
    \normalfont B\kern-0.5em{\scshape i\kern-0.25em b}\kern-0.8em\TeX}}}

\copyrightyear{2022}
\acmYear{2022}
\setcopyright{acmcopyright}\acmConference[SERP4IoT'22]{4th International
Workshop on Software Engineering Research and Practice for the IoT}{May 19,
2022}{Pittsburgh, PA, USA}
\acmBooktitle{4th International Workshop on Software Engineering Research and
Practice for the IoT (SERP4IoT'22), May 19, 2022, Pittsburgh, PA, USA}
\acmPrice{15.00}
\acmDOI{10.1145/3528227.3528567}
\acmISBN{978-1-4503-9332-4/22/05}

\newcommand{\etal}{\emph{et~al}. }
\newcommand{\ie}{\emph{i}.\emph{e}., }
\newcommand{\eg}{\emph{e}.\emph{g}., }
\newcommand{\cf}{\emph{cf}. }

\usepackage{booktabs}
\usepackage{hyperref}
\usepackage{subcaption}
\usepackage{makecell}
\newcommand{\xbl}{\textbf{BL}}
\newcommand{\xfi}{\textbf{FI}}
\newcommand{\xsh}{\textbf{SH}}
\newcommand{\xfish}{\textbf{FI}$\times$\textbf{SH}}

\usepackage{listings}
\definecolor{eclipseStrings}{RGB}{42,0.0,255}
\definecolor{eclipseKeywords}{RGB}{127,0,85}
\colorlet{numb}{magenta!60!black}
\lstdefinelanguage{json}{
    basicstyle=\footnotesize\ttfamily,
    commentstyle=\color{eclipseStrings}, 
    stringstyle=\color{eclipseKeywords}, 
    numbers=left,
    numberstyle=\scriptsize,
    stepnumber=1,
    numbersep=8pt,
    showstringspaces=false,
    breaklines=true,
    string=[s]{"}{"},
    comment=[l]{:\ "},
    morecomment=[l]{:"},
    literate=
        *{0}{{{\color{numb}0}}}{1}
         {1}{{{\color{numb}1}}}{1}
         {2}{{{\color{numb}2}}}{1}
         {3}{{{\color{numb}3}}}{1}
         {4}{{{\color{numb}4}}}{1}
         {5}{{{\color{numb}5}}}{1}
         {6}{{{\color{numb}6}}}{1}
         {7}{{{\color{numb}7}}}{1}
         {8}{{{\color{numb}8}}}{1}
         {9}{{{\color{numb}9}}}{1}
}
\usepackage{float}
\newfloat{listing}{htbp}{lop}
\floatname{listing}{Listing}

\begin{document}

\title{Evaluation of IoT Self-healing Mechanisms using  Fault-Injection in Message Brokers}


\author{Miguel Duarte}
\email{up201606298@fe.up.pt}
\affiliation{%
  \institution{Faculty of Engineering,\\University of Porto}
  \city{Porto}
  \country{Portugal}
}

\author{João Pedro Dias}
\email{jpmdias@fe.up.pt}
\affiliation{%
\institution{BUILT CoLAB and}
  \institution{Faculty of Engineering,\\University of Porto}
  \city{Porto}
  \country{Portugal}
}

\author{Hugo Sereno Ferreira}
\email{hugosf@fe.up.pt}
\affiliation{%
  \institution{INESC TEC and}
  \institution{Faculty of Engineering,\\University of Porto}
  \city{Porto}
  \country{Portugal}
}
\author{André Restivo}
\email{arestivo@fe.up.pt}
\affiliation{%
    \institution{LIACC and}
  \institution{Faculty of Engineering,\\University of Porto}
  \city{Porto}
  \country{Portugal}
}

\renewcommand{\shortauthors}{Duarte, et al.}

\begin{abstract}
The widespread use of Internet-of-Things (IoT) across different application domains leads to an increased concern regarding their dependability, especially as the number of potentially mission-critical systems becomes considerable. Fault-tolerance has been used to reduce the impact of faults in systems, and their adoption in IoT is becoming a necessity. This work focuses on how to exercise fault-tolerance mechanisms by deliberately provoking its malfunction. We start by describing a proof-of-concept fault-injection add-on to a commonly used publish/subscribe broker. We then present several experiments mimicking real-world IoT scenarios, focusing on injecting faults in systems with (and without) active self-healing mechanisms and comparing their behavior to the baseline without faults. We observe evidence that fault-injection can be used to (a)~exercise in-place fault-tolerance apparatus, and (b)~detect when these mechanisms are not performing nominally, providing insights into enhancing in-place fault-tolerance techniques.
\end{abstract}



\maketitle

\section{Introduction}

Internet-of-Things (IoT) is being largely adopted, being ubiquitous across application domains, connecting the physical and virtual realms to provide services that improve the quality-of-life and business processes~\cite{RoyWantBillNSchilit2017}. IoT devices are typically constrained in both computational power and energy, highly distributed (both logically and geographically), and heterogeneous (\eg different manufacturers and competing standards). The nature of these systems drives the creation of communication protocols that are lightweight and thus compatible with the computational and energy constraints of these systems. Among those protocols, MQTT has been largely adopted as a lightweight TCP-based Machine-to-Machine (M2M) and IoT connectivity protocol
~\cite{hillar2017mqtt}. MQTT leverages a publish/subscribe pattern, in which a middleware broker guarantees the delivery of messages from publisher entities (\ie entities that publish messages in a given topic) to one or more subscriber entities (\ie entities that are subscribed to a given topic)~\cite{hillar2017mqtt}.

IoT dissemination required end-users to program their own systems, thus leading to the proliferation of low-code development platforms such as Node-RED which has a visual programming language to mashup together hardware devices, APIs, and online services~\cite{noderedwebsite,torres2020real,Ihirwe2020,silva2021review}. However, most of these low-code development environments lack of verification and validation mechanisms~\cite{Diasbrief18}, and do not provide fault-tolerance mechanisms or suggest how to improve the dependability of these systems. This is expected, as most IoT systems disregard these concerns or achieve redundancy only by device/service replication (with additional costs and management complexity)~\cite{Ray2016,Cheng2017,Javed18}. This is a mostly direct result of the complexity of creating and using fault-tolerance mechanisms in IoT systems, as pointed by Javed \etal, \emph{``building a fault-tolerant system for IoT is a complex task, mainly because of the extremely large variety of edge devices, data computing technologies, networks, and other resources that may be involved in the development process''}~\cite{Javed18}. 

The use of self-healing\cite{Ganek2003} --- the ability of a system to automatically detect, diagnose and repair system defects at both hardware and software levels --- to attain fault-tolerance on IoT systems has been suggested by several authors~\cite{Angarita2015,Dundar16,Seiger17,ramadas2017patterns,Aktas2019,seeger2019optimally}. In previous work~\cite{europlop19} a set of patterns to achieve fault-tolerance in IoT systems by adding self-healing mechanisms was introduced, along with a reference implementation in Node-RED. This reference implementation, so-called SHEN\footnote{SHEN is available on GitHub: \texttt{node-red-contrib-self-healing}, \url{https://github.com/jpdias/node-red-contrib-self-healing}.}, consists of a set of self-healing add-on \textit{nodes} to the Node-RED visual programming language that can be used to improve the visual \textit{flows} with error detection and system health recovery/maintenance mechanisms~\cite{patternshealing,selfheal20,icse2021}. 

One way of ensuring that a self-healing/fault-tolerance mechanisms work as intended, is to actually exercise them. In the research field of fault-tolerance, fault-injection has been used as a technique to deliberately cause errors and failures in systems by introducing faults and then observing how it behaves and recovers from them; a now common practice in Chaos Engineering~\cite{basirinetflix2016}.

For the purposes of this work, we assume that the IoT system under study uses MQTT as the communication substrate, thus requiring a message broker to manage all the combinations, and that there is no redundant broker units. Thus, we can instrument an MQTT broker to inject faults in the messages as they are exchanged in the broker. And, as stated in the \textit{fault model} introduced by Esposito~\cite{Esposito2013}, we are able to introduce several types of faults at the network level (more specifically, at the message level) namely: (1)~\textit{omission}, (2)~\textit{corruption}, (3)~\textit{reordering}, (4)~\textit{duplication}, and (5)~\textit{delay}. With an instrumented broker capable of injecting faults on-demand in a running IoT system with self-healing mechanisms, we can: (a)~exercise the in-place fault-tolerance mechanisms, and (b)~know when these mechanisms are not working correctly, thus finding improvement targets. We proceed to collect empirical evidence that supports these claims by defining two experimental scenarios and a total of six experiments.

The main contributions of this paper are twofold: (1)~an instrumentable MQTT broker that allows fault-injection in IoT systems either using pre-defined operators or user-defined ones, and (2)~a validation of the self-healing extensions for Node-RED~\cite{selfheal20,icse2021}.

The remaining of the paper is structured as follows: Section~\ref{sec:related} presents some related work and Section~\ref{sec:broker} introduces fault-injector proof-of-concept, and Section~\ref{sec:preliminaries} details the experimental setup. Section~\ref{sec:experiments} presents the carried experiments and the obtained results. Finally, Section~\ref{sec:discussion} presents an overall discussion of the experiments and results, Section~\ref{sec:threats} discusses some of the threats to the validity of the experiments carried in this work, and Section~\ref{sec:conclusion} gives some closing remarks and points to future work.

\section{Related Work}
\label{sec:related}

Although most literature regarding IoT and fault-injection focuses on hardware faults via physical interaction with devices \cite{kazemi2020_review_hardware_fault_injection_iot,zabib2017_iot_laser_fault_injection, holler2015_qemu_fault_injection_hardware_fault_attacks} rather than interfering with the application layer logic (including communication channels and middleware components), the usage of software techniques to exercise in-place fault-tolerance mechanisms can be traced to as early as 1993~\cite{Arlat1993}. More recently, particularly in cloud computing, Chaos Engineering became an umbrella term to techniques that inject, observe and collect information concerning the impact of faults on a running system~\cite{basirinetflix2016,Boldt18,cotroneo2020fault}.

Looker and Xu~\cite{Looker03} presented an approach to carry fault-injection in the Open Grid Services Architecture (OGSA) middleware, which ensures the exchange of messages across services (\eg cloud). Jacques-Silva~\etal~\cite{JacquesSilva11} suggested the use of partial fault-tolerance (PFT) techniques to improve the dependability of stream processing applications, given that faults in computational nodes or stream operators can provoke massive data losses due to the high data exchange rates of these systems. Esposito~\cite{Esposito2013} surveyed research between 1995 and 2013 on fault-injection in the context of communication software; they found most of them focused on faults at the message-level and other networking failures (38.8\%), with the remaining on process crashes (12.24\%), application-level faults (18.4\%), memory corruptions (8.2\%), and others (22.36\%). They also introduce a fault-injection approach for publish/subscribe systems that encompass most of the faults presented in previous works. Yoneyama~\etal~\cite{yoneyama2019_fault_injection_iot_protocols} performed model-based fault-injection on MQTT, by mimicking unstable network environments via simulated network errors (\eg connection lost) and delays. Han~\etal~\cite{HanTrack19} introduced TRAK as a message-agnostic testing tool for injecting delays and provoke higher packet loss in Apache Kafka~\cite{garg2013apache}, thus asserting its QoS capabilities. Zaiter~\etal~\cite{Zaiter20} presented a distributed fault-tolerance approach for e-health systems based on the exchange of messages between LACs (Local Controlling Agents), reducing the reliance on one global agent controller (GAC). Briland~\etal~\cite{Briland2021} explores faults where third-parties inject fabricated data and expect it to modify the system's behaviour over time; they propose a Domain-Specific Language (DSL) to generate altered data that can then be  injected into the system to observe its behaviour. 

In summary, previous software-based fault-injection literature mostly explores faults at the communication/protocol level~\cite{cvartho2013_modbat}, with few tackling domain-specific behaviors (\eg modifying sensor readings). Most also rely on fault-injection agents as new system's components, with a single work preferring to modify the middleware~\cite{HanTrack19}). This limits their usage in IoT due to the computational constraints of most entities. Finally, very few works use fault-injection to evaluate the behaviour of in-place fault-tolerance mechanisms~\cite{yoneyama2019_fault_injection_iot_protocols,JacquesSilva11,HanTrack19,Zaiter20}. This paper improves on existing work by (1)~creating faults by semantically changing messages passed between different parts of the system, (2)~providing a DSL comprised of reactive operators\footnote{Inspired by the ReactiveX operators~\cite{maglie2016reactivex}, \url{http://reactivex.io}}, (3)~modifying a common middleware to target any MQTT-based system, and (4)~designed to support in-place evaluation of fault-tolerance mechanisms.


\section{Instrumented MQTT Broker}
\label{sec:broker}

To execute the experiments, we modified an commonly used open-source  MQTT broker\footnote{AEDES MQTT broker, \url{https://github.com/moscajs/aedes}} to inject faults into IoT systems. We choose fault-injection at the message broker level due to the fact the broker is a point of convergence of most IoT systems, mostly independent of the message publish/subscribe complexity, thus reducing the impact of the heterogeneity of the IoT system in the fault-injection strategies. The modifications were done to the broker allowed to use it as a proxy to intercept and modify messages before being published to a specific topic. The modified broker is available on GitHub\footnote{\textit{Instrumentable-Aedes}, \url{https://github.com/SIGNEXT/instrumentable-aedes}};  more details on its inner workings can be found in~\cite{miguelduartethesis}.

Each fault-injection rule consists of a \texttt{topic} (where the rule will be applied), and an array of \texttt{operators} each one transforming the incoming message and passing it to the next one (as in a \emph{pipes and filters} architecture). Each rule can also have a \texttt{startAfter} and \texttt{stopAfter} fields that define the number of messages before the faults start and stop being injected. The implemented operators in this proof-of-concept are the following: (1)~\texttt{map}, which takes a function as an argument (\eg multiply by two) which is then applied to the message, (2)~\texttt{randomDelay} that delays the publication of a message, (3)~\texttt{buffer} which captures all the messages until a time interval and/or number of messages is reached, publishing them all at once, and (4)~\texttt{randomDrop}, which randomly drops some of the messages according to a specified probability.  While all operators could be implemented through the \texttt{map} one, others are provided for ease the configuration of fault-injection, being the \texttt{map} operator useful for creating additional faults and transformations on-demand.

These operators were chosen to be implemented since they can be used to mimic some common faults in sensors and IoT systems as a whole~\cite{Ni2009,patternshealing}, faults which can result from the malfunctioning of a given device (\eg out-of-spec sensor readings can be mimic with a \texttt{map} that multiplies the readings by a random number), network issues (\eg  message loss with \texttt{randomDrop} and lag with \texttt{randomDelay}), or a combination of these factors (and others). These concrete operators were also selected since they can be used to validate the existent Node-RED self-healing \textit{nodes}.  Given the aforementioned operators, we can inject all the faults presented in the \textit{fault model} of Esposito~\cite{Esposito2013}.

\section{Preliminaries}
\label{sec:preliminaries}


The possible combinations of the system with and without self-healing or fault-injection result in four variations of the system under test (SUT). We called these \xbl{} (baseline), self-healing (\xsh{}), fault-injection (\xfi), and self-healing with fault-injection (\xfish{}).

If the fault-injection and self-healing mechanisms are working correctly we expect that (1)~the behavior of \xsh{} approximates \xbl{}, as no fault-injection is performed in either system and self-healing mechanisms should have a low impact in a nominal system; (2)~the behavior of \xfi{} is very different from \xbl{}, since the base system, without self-healing components, should not be able to recover from injected faults, provided the fault is enough to deviate it from nominal operation; and (3)~the behavior of \xsh{} is similar to that of \xfish{}, showing that the self-healing mechanisms are able to bring a system with injected faults back into nominal behavior.

These assumptions are expected to hold since the self-healing mechanism for each one of the scenarios was selected in accordance with the type of sensing data, sensing data cadence, sensor maximum, and minimum reading values (\ie device specifications). Other aspects, including network latency and packet loss, are expected to have no impact since the experiments were run in a single machine. Further, the injected faults are directly related to the type and frequency of sensing data being exchanged; thus, it is expected that such errors would emerge in a real-world system. 

The experiments were done on a standard Linux laptop with Node-RED version 1.3.2, and the modified AEDES MQTT broker was run with NodeJS version 14.15.5. A replication package for these experiments is available on Zenodo~\cite{duarte_miguel_2021_5148566}. 

\section{Experiments and Results}
\label{sec:experiments}

Two test scenarios (\textbf{S1} and \textbf{S2}) were devised and several experiments were done for each scenario. Each one of these used a separate fault-injection configuration so that different operations were applied to the same dataset. For each experiment, messages were relayed to both the baseline (\xbl{}) and the corresponding self-healing (\xsh{}) flow simultaneously to ensure that both systems received the same input and that their outputs (\ie alarm level) could be directly compared. This also ensured that the fault-injection operators, which rely on the MQTT broker's random number generation, do not produce different results when compared to the baseline.

To remain as close as possible to a real-world system, we used a real-world dataset\footnote{Dataset available at \url{https://archive.ics.uci.edu/ml/datasets/Air+Quality} and \url{http://archive.ics.uci.edu/ml/machine-learning-databases/00360/}.} as sensing data. The dataset used contained $\text{NO}_{x}(\text{GT})$ readings from a device \textit{``located on the field in a significantly polluted area, at road level, within an Italian city''}~\cite{DeVito2008_nox_dataset}. The sensing data was \textit{replayed} --- \ie each data point was \textit{emitted} using a Python script with a reduced time interval between readings (in the dataset one data point was collected per hour) --- reducing the time required for each experiment to run, thus allowing to increase the number of experiments and refinements. Since the dataset does not provide the concrete specification of the sensor's reading range of values, we considered a widely available sensor\footnote{\url{https://www.aeroqual.com/product/nitrogen-dioxide-sensor-0-1ppm}.} as a reference, which has a hardware range of 0-1 ppm (0-1000 \emph{ppb}) and a minimum detection limit of 0.005 ppm (5 \emph{ppb}). Thus, only values ranging from 5 to 1000 \emph{ppb} should be considered valid readings. We considered each reading should trigger an alarm according to three concern levels\footnote{While the \textit{EPA National Ambient Air Quality Standards list 0.053ppm as the avg. 24-hour limit for NO2 in outdoor air}\cite{Al-Sultan2019_NO2}, for validation purposes we will consider these limits for each data point and not the 24-hour average value.}: \textit{off} (0), \textit{warn} (1), and \textit{danger} (2). A threshold of 53 \emph{ppb} was considered as a \textit{warn}ing value, and 212 \emph{ppb} as a \textit{danger} value. 

Each experiment replays a total of 120 messages from the used dataset, and faults are injected to messages 10 thru 110 --- this allows the system to gain stability before entering a degradation state and can also resume normality after fault-injection stops.  In order to collect insights on the behavior of the SUT, all the messages flowing in the different MQTT topics are monitored and logged during the course of the experiment. All the experiments use the same \xbl{}, with three $\text{NO}_{x}$ sensors used to trigger an alarm according to the defined thresholds. The corresponding \xbl{} system, implemented as a Node-RED flow, parses the reading received from the sensors and sends the respective alarm level as output, filtered by a report-by-exception node to ensure that values are only emitted when the current alarm level changes. The Node-RED \textit{flows} with self-healing capabilities used in the following scenarios use a subset of the SHEN \textit{nodes}, as presented in previous work~\cite{icse2021}.

\subsection{Sensor Readings Issues (S1)}

Here (\textbf{S1}) we performed four experiments, each with different types of fault-injection operators applied to the sensor messages passing through the MQTT broker (\ie simulating sensor malfunctions).

A Node-RED flow, with self-healing mechanisms, was developed to deal with these issues (\xsh{}). This system expands upon \xbl{} by introducing self-healing capabilities via \texttt{SHEN} nodes. It filters extraneous messages that are outside the expected operating range, compensates for missing values after a certain timeout, joins messages so that they are considered in groups of 3, considers the majority of values with a minimum consensus of 2 (with a 25\% difference margin), and compensates for readings for which there is no majority with a mean of the previous readings, besides the basic functionality implemented by \xbl{}. The \texttt{join} and \texttt{compensate} \textit{nodes} are configured with a timeout of 6 seconds to have a margin of 1 second in relation to the readings' periodicity (5 seconds).

\subsubsection{Experiment S1E1}

In this experiment no fault injection was performed, with only the baseline system (\xbl{}) and the system with added self-healing mechanisms (\xsh{}) being considered. This allows us to compare the behavior of \xbl{} with \xsh{} in normal operation, and creates a base of comparison for the remaining experiments (\ie experiments with fault-injection). This also provides us a behavioral profile of \xbl{} when compared with \xsh{}, giving us insights on how self-healing mechanisms' operate with no added entropy.

We expect the SUT to remain stable during this experiment, outputting the expected alarm levels for the sensor readings' thresholds. Despite the expected similarity in behaviour, it is expected that \xsh{}'s alarm level output will be more stable than that of \xbl{}. This is due to the latter not implementing any type of consensus or majority voting and instead simply using the received values directly as a stream.

\begin{figure}[h]
		\centering
		\includegraphics[width=\linewidth]{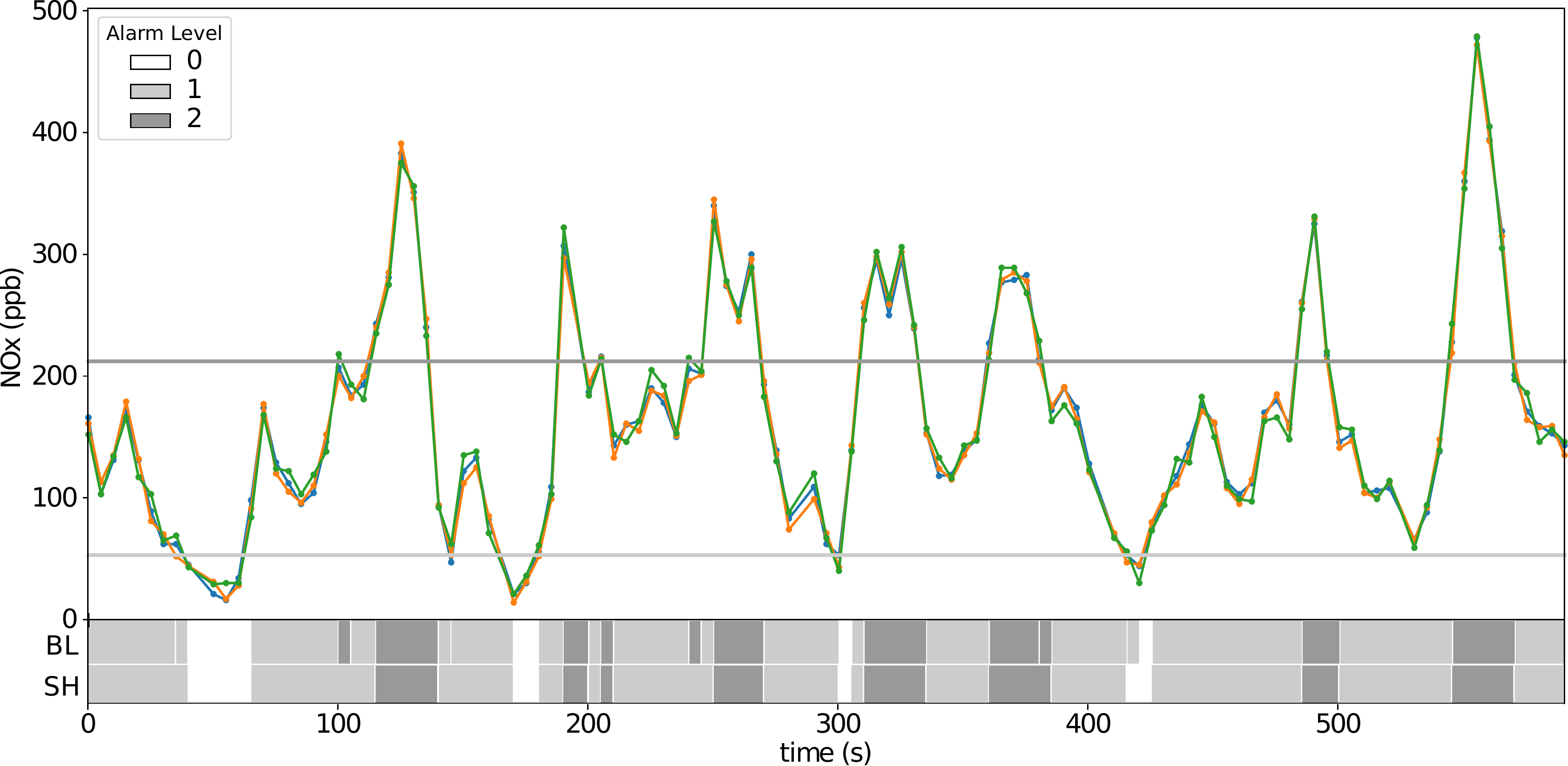}
		
	\caption[$\text{NO}_{x}$ concentration and alarm status for S1E1.]{$\text{NO}_{x}$ concentration and alarm status for S1E1.}
	\label{fig:validation_s1-e1}
\end{figure}

Fig.~\ref{fig:validation_s1-e1} shows the experiment results for \xbl{} and \xsh{}. Despite the alarm output (represented as shaded areas near the horizontal axis) being very similar for both experiment outputs, stability is higher for \xsh{}. This can be observed by the lack of fast alarm state changes for borderline values for \xsh{}, which occur several times for \xbl{} (\eg around 35 to 40 seconds into the experiment or around 415 to 420 seconds). The cause of this is likely to be \xbl{}'s lack of consensus mechanism, given that this system instead simply considers the most recent reading in order to determine the alarm state. When the three sensors report in quick succession, if the values of their readings are near the alarm level thresholds, fluctuations in the alarm level are expected.

The output similarity is confirmed by the alarm level overlap percentage between these two outputs, which is $97.3\%$. This is also a good sanity check to confirm that the addition of self-healing capabilities to the base system does not significantly change the alarm output status, which means that comparisons for \xsh{} between \textbf{S1E1}, and further experiments, will be meaningful in validating self-healing recovery of injected faults.

\begin{table}[h]
	\centering
	\caption[Count of alarm level state transitions for S1 experiments.]{Count of alarm level state transitions for S1E1-4.}
	\label{tab:s1_alarm_state_transitions}
	\begin{tabular}{lrrrrrrrr}
		\toprule
		&    \multicolumn{2}{c}{S1E1} &    \multicolumn{2}{c}{S1E2}  & \multicolumn{2}{c}{S1E3}  &    \multicolumn{2}{c}{S1E4} \\ \midrule
		\multicolumn{1}{c}{}                    & \xbl{} & \xsh{} & \xbl{} & \xsh{} & \xbl{} & \xsh{}& \xbl{} & \xsh{} \\ \midrule
		\multicolumn{1}{l}{\textit{Off} (0)}    & 8                      & 4     & 10                     & 5      & 8                      & 4       & 8                      & 4          \\ 
		\multicolumn{1}{l}{\textit{Warn} (1)}   & 20                     & 13   & 68                     & 14     & 26                     & 13         & 20                     & 13        \\ 
		\multicolumn{1}{l}{\textit{Danger} (2)} & 11                     & 8      & 70                     & 8 & 17                     & 8          & 11                     & 8           \\ \midrule
		\multicolumn{1}{l}{\textbf{Total}}      & 39                     & 25     & 148                    & 27  & 51                     & 25         & 39                     & 25           \\
		\bottomrule
	\end{tabular}
\end{table}

Table \ref{tab:s1_alarm_state_transitions} supports the previous claim --- that \xsh{} provides an improvement in system stability in comparison to \xbl{} --- due to the lower number of alarm state transitions. Additionally, this experiment presents evidence that both \xbl{} and \xsh{} correctly implement the expected core functionality (triggering the different alarm levels for different sensor reading thresholds), given that the alarm level at a given point in time corresponds to the sensor readings' distribution along the thresholds (represented in the mentioned figures by the horizontal lines).


\subsubsection{Experiment S1E2}

Considering the baseline (\xbl{}) and the self-healing (\xsh{}) systems (which S1E1 shows to be similar in normal operation), we proceed to inject faults on both, obtaining systems \xfi{} (corresponding to the injection of faults in \xbl{}) and \xfish{} (corresponding to the injection of faults in \xsh{}).

The fault being injected corresponds to an erroneous \textit{Sensor 3}'s reading. As a result, this sensor's readings are altered to be stuck at the upper operating bound (1000 \emph{ppb}). This experiment simulates a fault in which a sensor malfunctions by continuously emitting readings in its top operating bound.

We expect that this fault-injection will provoke an inconsistent output in \xfi{}, especially if the third sensor is frequently the last to emit its reading, even if only by a slight delay. Due to relying on a majority of at least two values to decide on the alarm level to emit, we expect that the self-healing mechanism will be able to deal with the faults injected in \xfish{}.

\begin{figure}[h]
	\centering
    \includegraphics[width=\linewidth]{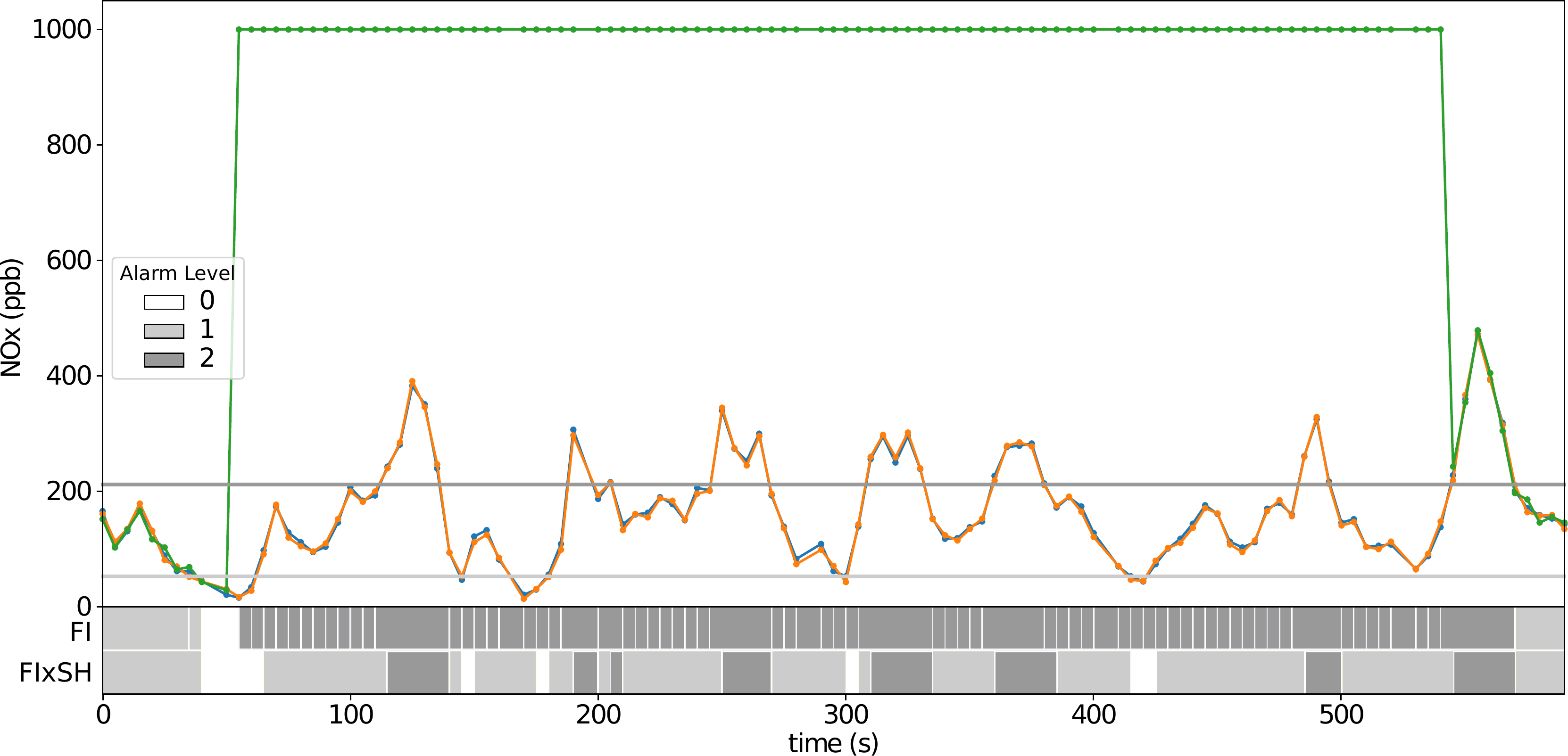}
	\caption[$\text{NO}_{x}$ concentration and alarm status for \textbf{S1E2}]{$\text{NO}_{x}$ concentration and alarm status for \textbf{S1E2}.}
	\label{fig:validation_s1-e2}
\end{figure}

Fig.~\ref{fig:validation_s1-e2} shows the experiment results for \xfi{} and \xfish{}. The faults injected (\xfi{}) disrupt the normal function of the system, resulting in constant alternation between alarm states, spending most of the experiment's time in the highest alarm level. Meanwhile, \xfish{} successfully recovers from the injected faults, having a near-perfect performance in comparison to this system's output for \textbf{S1E1}.

These statements are supported by the overlap in alarm levels between \xfish{} and \xsh{}, with a near-perfect overlap of 98.1\%, and \xfi{} and \xbl{}, with a  much lower overlap percentage of 40.0\%. Table \ref{tab:s1_alarm_state_transitions} also illustrate these conclusions, in which \xfi{} is much more unstable in comparison to \xbl{}, being the total number of state transitions for this experiment 148, while there were only 39 state transitions for the base experiment. Furthermore, the number of state transitions with self-healing has increased only marginally, going from 25 in the base experiment (\xsh{}) to 27 in this experiment (\xfish{}).

Therefore, \textbf{S1E2} demonstrates that the original system cannot handle sensor \textit{stuck-at} issues since the behavior of \xfi{} is considerably affected, which validates that the performed fault-injection was meaningful enough to disturb the system's regular operation. On the other hand, \xfish{} can recover from the injected faults, having a remarkably similar behavior to \xsh{}. This happens since the \textit{stuck-at} fault only affects one sensor, and with the usage of the \texttt{replication-voter} node, this reading will be discarded and the other two sensors' values will be considered instead, resulting in a system that operates similarly to the \xsh{}.

\subsubsection{Experiment S1E3}

In this experiment, we injected faults in \xbl{} and \xsh{} to obtain \xfi{} and \xfish{} by multiplying 40\% of the readings done by \textit{Sensor 3} by a random factor in the range $[0.2, 2.2]$, simulating \textit{spikes} in sensor readings. The factor is randomized for each \textit{spike} occurrence\footnote{This fault is common for sensing devices when they are running out of battery~\cite{Ni2009}.}. 

We expect that \xfi{} may output incorrect alarm values (in comparison to \xbl{}), especially when the altered values switch between alarm level thresholds. On the other hand, \xfish{} should be able to handle the \textit{spikes} since that, even if one of the three sensors outputs a value considerably different from the others, it will be discarded and the other two sensors' values will be considered instead --- due to the usage of the \texttt{replication-voter} node.

\begin{figure}[h]
		\centering
		\includegraphics[width=\linewidth]{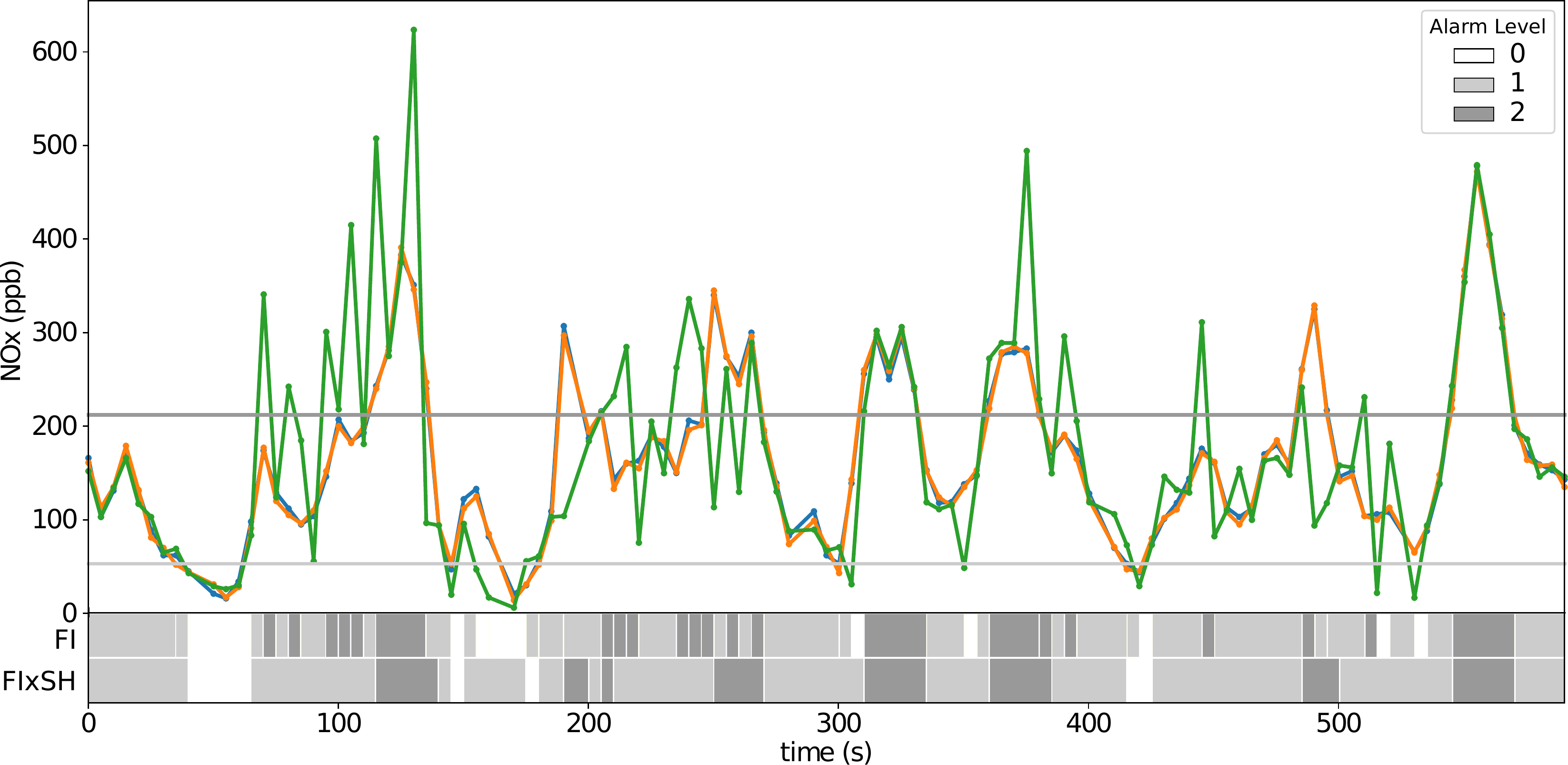}
		
	\caption[$\text{NO}_{x}$ concentration and alarm status for S1E3.]{$\text{NO}_{x}$ concentration and alarm status for \textbf{S1E3}.}
	\label{fig:validation_s1-e3}
\end{figure}

Fig.~\ref{fig:validation_s1-e3} shows the experiment results for \xfi{} and \xfish{}. \xfi{} has had a good performance in the presence of the \textit{spikes} (both when increasing and decreasing the read value), but there were still several situations in which the sensor reading \textit{spike} caused the output alarm level to differ from the expected value in \xbl{}. \xfish{} has held up to the defined expectations, handling almost all the injected faults and operating similarly to \xsh{}.

These statements are supported by the overlap in alarm levels between \xfish{} and \xsh{}, with a near-perfect overlap of 97.4\%, and \xfi{} and \xbl{}, with a lower overlap percentage of 76.3\%, showcasing the disruption provoked by the injected \textit{spikes}.

Table \ref{tab:s1_alarm_state_transitions} supports the role of the self-healing mechanisms. Despite the difference not being as remarkable as that of the overlap percentages, it is of note to mention that \xfish{} has the exact same number of alarm level state transitions of \xsh{} while the number of state transitions has increased for \xfi{} when compared with \xbl{}.

Despite this experiment not causing a variation as significant as that of the behavior of \xfi{} in \textbf{S1E2}, we were still able to observe a mismatch between the behavior of this system between the base case and this experiment, even if to a lesser extent. This shows that for the system under study, the \textit{spikes} faults are less concerning when compared with the \textit{stuck-at} ones injected in \textbf{S1E2}. Nevertheless, due to the decline in the overlap percentage for \xfi{} in comparison with \xbl{}, we can conclude that the faults injected were significant enough to affect the system's correct functioning.

Since \xfish{} behaved similarly to \xsh{}, we can confirm that for this experiment the presence of self-healing capabilities are beneficial for the system's correct operation, thus improving its resilience.

\subsubsection{Experiment S1E4}

In this experiment, we injected faults in \textit{Sensor 3} so that it has a 20\% chance of losing messages\footnote{This fault may occur when a sensor is disconnected, has an intermittent power supply, or the network is unstable~\cite{Ni2009}.}. The system does not receive any of the lost messages, as these are suppressed before leaving the message broker. 

We expect that \xfi{} may report erroneous alarm values (compared to the base experiment, \textbf{S1E1}), especially when the missing values are in proximity to the alarm thresholds. \xfish{} should be able to handle the injected faults by compensating the missing values by replaying the last message in the expected time interval.

\begin{figure}[h]
		\centering
		\includegraphics[width=\linewidth]{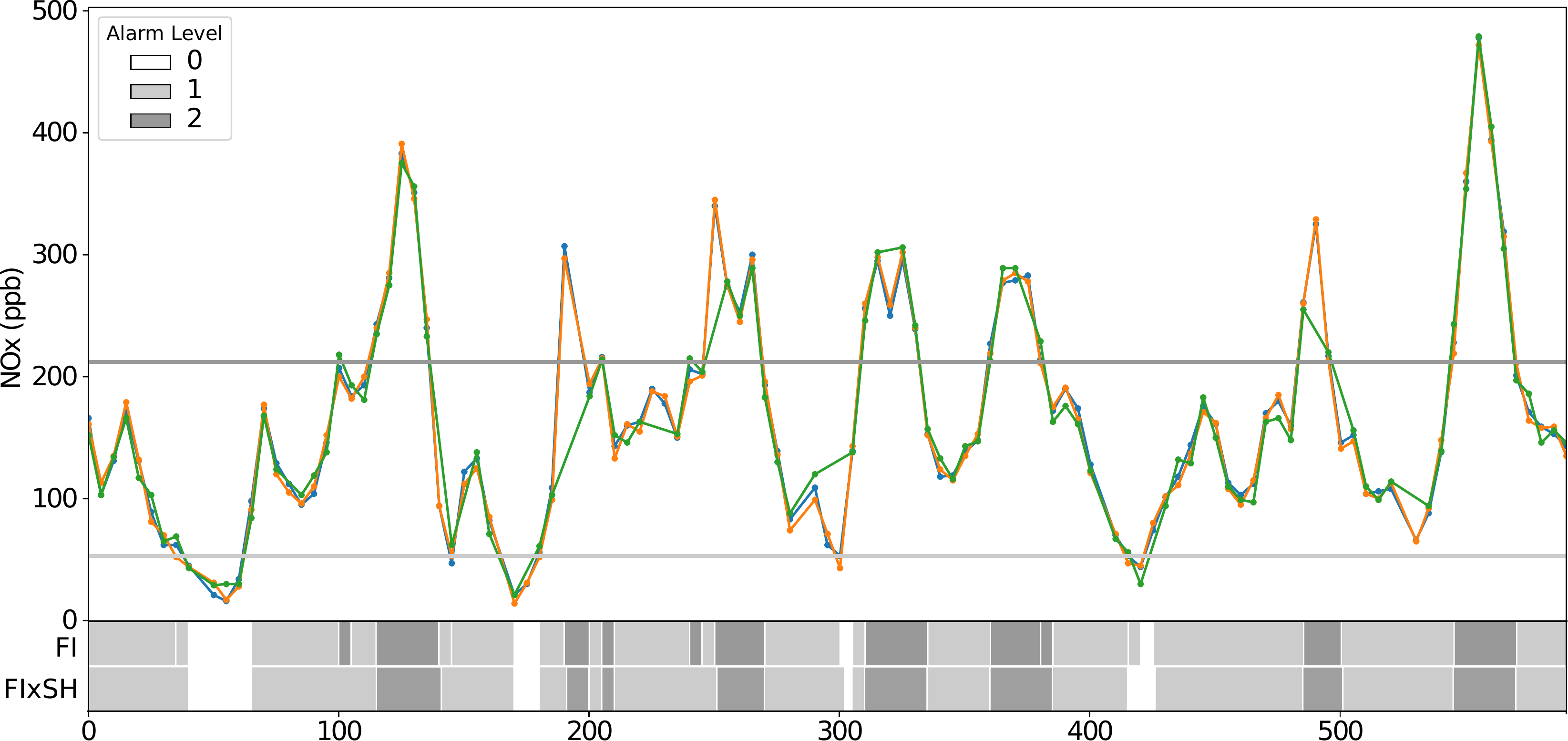}
		
	\caption[$\text{NO}_{x}$ concentration and alarm status with self-healing for S1E4.]{$\text{NO}_{x}$ concentration and alarm status for \textbf{S1E4}.}
	\label{fig:validation_s1-e4}
\end{figure}

Fig.~\ref{fig:validation_s1-e4} shows the experiment results for \xfi{} and \xfish{}. \xfi{} is capable of handling the loss of some readings, thus the alarm output is quite similar to \xbl{}. \xfish{} is also able to handle the loss of readings, similarly having almost the same behavior as \xsh{}.

The similarity in outputs when compared with \textbf{S1E1} is a direct result of the low probability of losing one reading. Also, since the values for different sensors in the original dataset are close to one another, even by increasing the probability of values being suppressed from one sensor would result in \xfi{} outputting the expected alarm values for most of the experiment's duration.

These statements are supported by the overlap in alarm levels between \xfish{} and \xsh{}, 98.7\%, and \xfi{} and \xbl{}, 99.8\%, showing that \xfi{} has a similar behavior to \xbl{}, and that \xfish{} has a similar behaviour to \xsh{}. The previous observations are also supported by the results in Table \ref{tab:s1_alarm_state_transitions}, which shows the number of state transitions for both systems to be identical to those that occurred in \textbf{S1E1}.

This experiment did not cause a significant enough deviation from the base experiment's behavior for \xfi{}, which is corroborated by the high overlap percentage with \xbl{} of 98.4\%, as well as the fact that the number and type of alarm state transitions are identical to those of \textbf{S1E1} (\cf Table \ref{tab:s1_alarm_state_transitions}). Thus, we conclude that the used dataset may not be the best candidate for this type of fault injection. A higher deviation in operation could possibly be observed if the fault-injection was done to more than one sensor at a time and with a higher probability of losing a message.

\subsection{Timing Issues (S2)}

In this scenario (\textbf{S2}), the Node-RED flow self-healing mechanisms used in \textbf{S1} was enriched with \textit{nodes} that detect and mitigate issues with timings (\eg readings frequency issues) by introducing \texttt{debounce} nodes which can filter out extraneous messages based on the expected timing of the system's regular messages. The \texttt{join} and \texttt{compensate} \textit{nodes} are configured with a timeout of 6 seconds to have a margin of 1 second in relation to the readings' periodicity (5 seconds). A total of two experiments were conducted for \textbf{S2}. 

\subsubsection{Experiment S2E1}

No faults are injected for this experiment. Similarly to \textbf{S1E1}, the purpose of this experiment is to confirm that the system's base functionality is correctly implemented for both \xbl{} and \xsh{}, as well as to provide a base experimental output with which to compare the behavior of the systems in following experiments. As with \textbf{S1E1}, we expected that the systems under observation remain stable during this experiment since there are no injected faults and that \xsh{}'s alarm level output will be more stable than that of \xbl{}. The results were similar to those of \textbf{S1E1} with a similarity of 97.4\%.

\subsubsection{Experiment S2E2}

\begin{figure}[h]
\begin{center}
	\includegraphics[width=.9\linewidth]{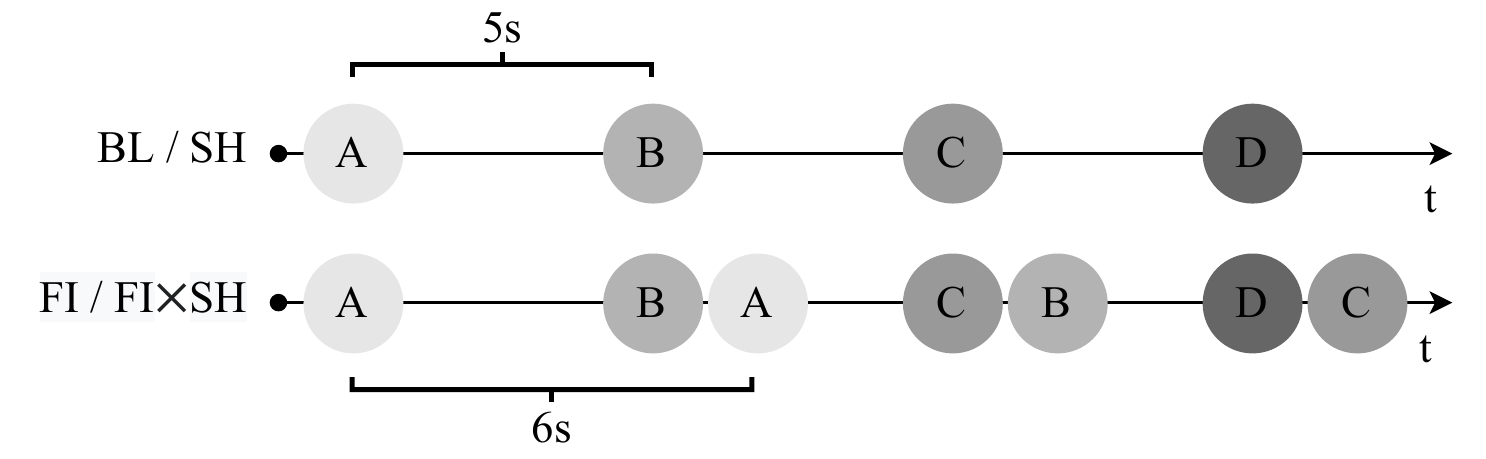}
	\caption[Marble diagram of messages for S2E2.]{Marble diagram of messages for S2E2. The top diagram depicts the regular flow of messages, while the bottom diagram shows the messages after the fault injection.}
	\label{fig:marble_s2e2}
\end{center}
\end{figure}

In this experiment, to introduce additional noise into the system, each message for \textit{Sensor 3} is repeated after 6 seconds, as depicted in Fig.~\ref{fig:marble_s2e2}. Since the periodicity of the system's readings in regular circumstances is of 5 seconds, the repeated message will be outputted in close proximity to the next reading.

We expect \xfi{} to have an output that is less stable than it was for \xbl{} due to the injected faults; this may be problematic for \xfi{} since it does not have any concept of message timing. On the other hand, \xfish{} should be able to cope with the injected faults since the \texttt{debounce} \textit{node} will filter out the additional messages that come out of the expected frequency, thus behaving similarly to \xsh{}.

\begin{figure}[h]
	\centering
	\includegraphics[width=\linewidth]{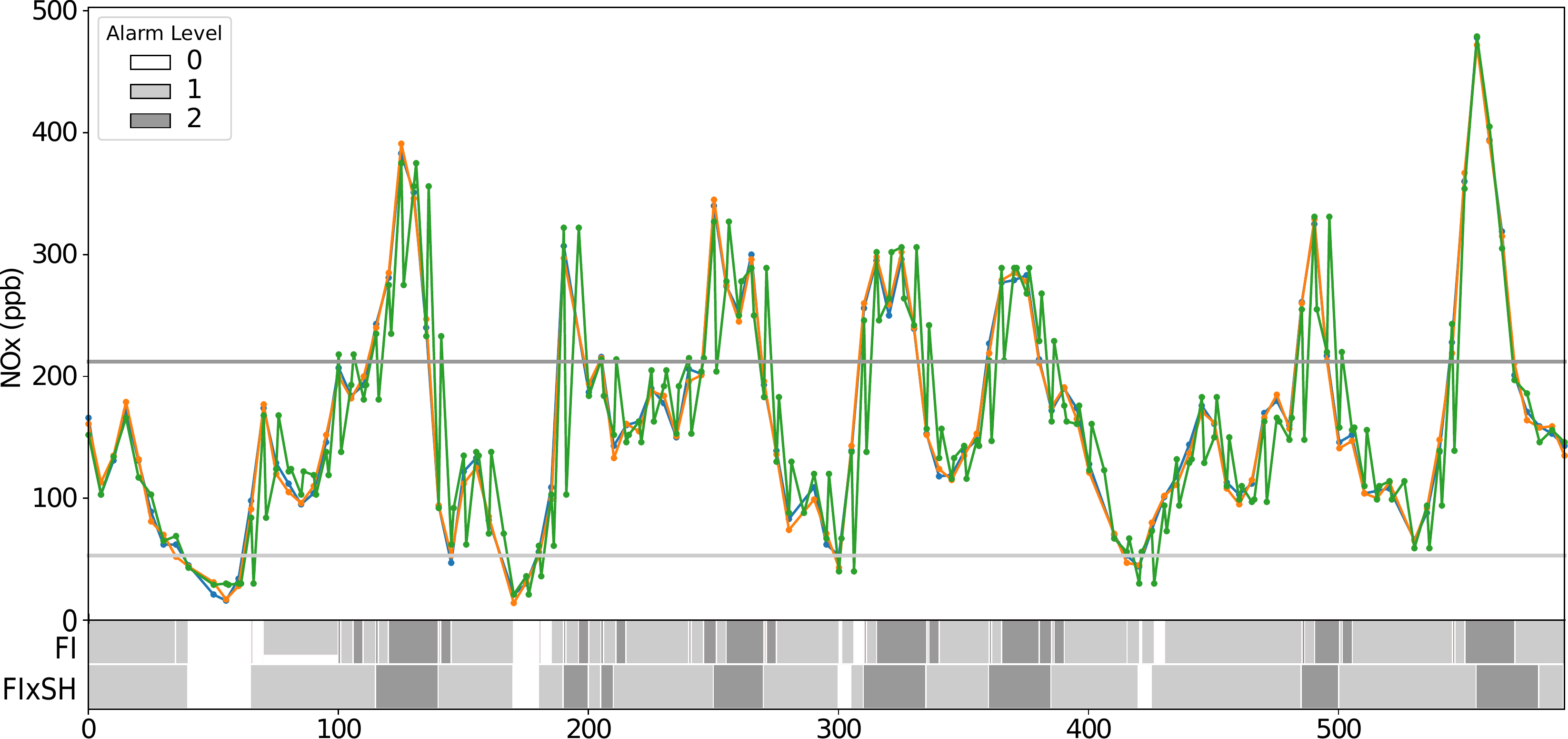}

\caption[$\text{NO}_{x}$ concentration and alarm status with self-healing for S2E2.]{$\text{NO}_{x}$ concentration and alarm status for \textbf{S2E2}.}
	\label{fig:validation_s2-e2}
\end{figure}

\xfi{} (Fig.~\ref{fig:validation_s2-e2}) performed significantly better than expected, despite the issues with messages near the alarm level thresholds, \ie repeating the previous message would cause the system to output the previous alarm level once again until it received the following message and went back to the expected state. This is caused by the fact that the periodicity of the sensor readings messages is quite high (\ie number of sensor readings per unit of time) and, whenever a fault is injected, it is not \textit{in effect} for a long duration. This can be checked by observing the number of alarm state transactions and their duration --- even with more state transactions, the time that the alarm stays at that given alert level is short --- resetting to a normal state when a new reading appears.

\begin{table}
	\centering
	\caption{Count of alarm level state transitions for S2E1-2.}
	\label{tab:s2e2_alarm_state_transitions}
	\begin{tabular}{lrrrr}
		\toprule
				&    \multicolumn{2}{c}{S2E1} &    \multicolumn{2}{c}{S2E2}  \\ \midrule
		          & \xfi{} & \xfish{} & \xfi{} & \xfish{} \\ \midrule
	\textit{Off} (0)  & 8                      & 4     & 12                     & 4                       \\ 
		\textit{Warn} (1)  & 20                     & 13   & 36                     & 13                      \\ 
		\textit{Danger} (2)  & 11                     & 8 & 25                     & 8                       \\ \midrule
	\textbf{Total}    & 39                     & 25     & 73                     & 25                      \\ \bottomrule
	\end{tabular}
\end{table}

As expected, \xfish{} was able to cope with the injected faults (Fig.~\ref{fig:validation_s2-e2}), having a near-perfect behavior in comparison to \xsh{}. These statements are supported by the overlap in alarm levels between \xfish{} and \xsh{}, with a near-perfect overlap of 95.7\%, and \xfi{} and \xbl{}, with a slightly lower overlap percentage of 83.4\%, showcasing the disruption provoked by the injected \textit{spikes}. Despite the high overlap percentages for \xfi{} with \xbl{}, Table \ref{tab:s2e2_alarm_state_transitions} shows that for \textbf{S2E2}, \xfi{} has output nearly two times the amount of alarm level state transitions in comparison to \textbf{S2E1}. 

\textbf{S2E2} shows that despite the introduction of faults in \xfi{} the difference shown by the overlap percentage to \xbl{} is minimal. Despite this, \xfish{} cope better with the injected faults, operating closer to \xsh{}. \xfi{} also performs worse than \xfish{} when taking into account the number of alarm level state transitions (\cf Table \ref{tab:s2e2_alarm_state_transitions}). And, while it is not possible to conclude if this experiment caused enough deviation from \xbl{} to \xfi{} by taking only in consideration the overlap percentage, the difference in the count of alarm level state transition for \xfi{} in comparison to \xbl{} provides evidence that the injected faults have caused issues on the baseline system which the self-healing system is able to sustain.

\section{Discussion}
\label{sec:discussion}

The fault-injection experiments \textbf{S1E1}, \textbf{S1E2}, \textbf{S1E3}, \textbf{S2E1}, and \textbf{S2E2} allowed us to observe that: (1) the self-healing systems (\xsh{}) do not deviate too much in behavior from the baseline system (\xbl{}); (2) the faults injected are consequential since there is a deviation on the baseline system in comparison to the base experiment when no fault is being injected; and (3) when the faults injected are consequential, the self-healing systems were able to recover from it, conforming with the normal service, and thus confirming that the self-healing mechanisms were being exercised and performing as expected.

More concretely, by analyzing the Table~\ref{tab:s1_alarm_state_transitions} and Table~\ref{tab:s2e2_alarm_state_transitions}, we can see the impact that fault-injection has in a system without any fault-tolerance mechanisms versus a system with self-healing capabilities. The number of times that the alarm changes state between its three alert levels is considerably higher in all experiments, with a clear impact in the experiments \textbf{S1E2}, \textbf{S1E3}, and \textbf{S2E2}, where the number of transitions was more than two times higher than the expected number of transitions. While the overlap percentages of the different experiments do not provide enough evidence to draw conclusions for most experiments, we can see that \textbf{S1E2} performs considerably better when self-healing mechanisms are present.

Additionally, \textbf{S1E4} allowed us to verify that it is paramount for the behavior of \xbl{} to be \emph{noticeably} different when faults are being injected (\xfi{}) in comparison to the regular operating circumstances. This factor made it so that we considered this experiment inconclusive due to the low entropy caused in the system. Nevertheless, it shows that it is necessary to find this stark difference in expected versus observed output for the baseline system to be sure if the self-healing components are doing any work at all since a naïve system would already be able to \textit{recover} from most injected faults.

It is also noticeable that for the created scenarios, due to the used dataset, some types of fault-injection did not result in much instability. This is due to the fact that all three sensors output readings with values in close to each other. As such, if one or even two sensors fail, it is likely that a naïve system (\eg \xbl{}) will still perform as expected, outputting the correct alarm levels for most cases even in the presence of faults (\xfi{}). This is an indicator that further validation should be performed with other types of datasets and systems, as well as different types of faults. 

\section{Threats to Validity}
\label{sec:threats}

The experiments presented in this work were carried out using a real-world dataset \textit{replayed} (\ie simulated) using a Python script. While all the messages were replayed, even if they contained erroneous or invalid readings, we cannot recreate any errors that could exist with the data if any pre-processing was done to the dataset before being put public available. This could be mitigated by using additional datasets of different IoT systems and carrying some extra, even if smaller, experiments in a real testbed. Further, while running the experiments in a simulated environment eases reproducibility, it has the downside of not covering sporadic faults that could occur in the physical testbed, \eg electromagnetic interference's and network disruptions.

An inadequate selection of the faults injected into the system also poses a threat to the validity of the experiments carried in this work since they have been hand-picked with prior knowledge of the fault-injection and self-healing capabilities. This can result in a bias in the selection, favoring issues that we know about and having more confidence that the proposed solution will handle correctly, instead of the ones that are mostly like to occur. While we attempt to mitigate this by using a real-world dataset, we have the bias of picking the faults injected and the self-healing logic.

Injecting faults at middleware (\ie message broker) also limits the range of possibilities. While we attempted to replicate some common sensor fault types~\cite{Ni2009}, other faults, including ones at the network level, were not covered by this study. Faults and implementation quirks of the underlying infrastructure --- \ie Python message emitter script, modified MQTT broker, Node-RED, and the self-healing extensions --- might also have influenced the outcome of the carried experiments, so they are a confounding variable. We believe this has been mitigated by careful analysis of the expected and actual results, though it is something of concern.

\section{Conclusions}
\label{sec:conclusion}

Ensuring the dependability of software systems has been the goal of most fault-tolerance research in the past years~\cite{Avizienis2001}. In IoT, ensuring security, reliability, and compliance is becoming a paramount concern due to the recent increase in mission-critical contexts. IoT fault-tolerance is considerably challenging due to several aspects, including (1)~high heterogeneity of devices, (2)~interaction and limitations of systems deployed in a \emph{physical} environment, (3)~field fragmentation, ranging from the high number of communication protocols to the different and competing standards, and (4)~intrinsic dependability on hardware that might simply fail~\cite{aly19}. 

Fault-injection becomes paramount to ensure that fault-tolerance mechanisms perform as they are expected when required. By instrumenting an MQTT broker, we enable the injection of faults at the message-passing level that allows to observe how well components deal with such faults. This provides a means to assert if the self-healing mechanisms configured in the system are sufficient to deal with them. The carried experiments showcase that the self-healing extensions do, indeed, work as expected, with the injected faults causing little to no impact on the delivery of normal service. 


As future improvements to the instrumented MQTT broker, we consider the following: (1)~simplify the fault-injection configuration by supporting more native language constructs (\eg arrow functions) and other configuration abstractions (\eg leverage visual notations), (2)~support wildcard topics as per the MQTT specification, and (3)~enable switching configuration at run-time instead of having to specify the configuration file when starting the broker. Regarding the experimental stage, it would be interesting to (1)~expand the scenarios with more experiments, including more extensive fault-injection pipelines; (2)~replicated the experiments using different datasets; (3)~extend the usage of self-healing mechanisms, especially the ones implemented as part of SHEN~\cite{icse2021}; and (4) explore decentralized IoT systems and orchestrators~\cite{pinto2018dynamic,silva2020visually}.

\begin{acks}
This work was partially funded by the Portuguese Foundation for Science and Technology (FCT), \emph{ref.} \texttt{SFRH/BD/144612/2019}.
\end{acks}

\bibliographystyle{ACM-Reference-Format}
\bibliography{myrefs}

\end{document}